\documentstyle{article}

\font\tenrm=cmr10

\font\elevenbf=cmbx10 scaled\magstep 1
\font\twelvebf=cmbx10 scaled\magstep 2
\font\elevenrm=cmr10 scaled\magstep 1
 2
 1

\renewenvironment{thebibliography}[1]
 { \elevenrm
   \begin{list}{\arabic{enumi}.}
    {\usecounter{enumi} \setlength{\parsep}{0pt}
     \setlength{\itemsep}{3pt} \settowidth{\labelwidth}{#1.}
     \sloppy
    }}{\end{list}}

\begin{document}
\vglue 2.5cm
{\twelvebf
Strange Skyrmions: status and observable predictions
\footnote{\tenrm\baselineskip=11pt
The work supported in part by the Russian Fund for Fundamental Researches,
grant 95-02-03868a}
\\}
\vglue 0.4cm
{\elevenrm 
Vladimir B. Kopeliovich \\}
{\tenrm Institute for Nuclear Research of the Russian Academy of
Sciences, Moscow 117312 \\}
 \vglue 0.1cm
\baselineskip=11pt
\tenrm
\vglue 0.3cm
{\bf Abstract} The chiral soliton approach (CSA) provides predictions of the
rich spectrum of baryonic states with different values of strangeness for any
baryon number $B$. In the sector with $B=1$ the well known octet and decuplet
of baryons are described within CSA, and some exotic sates are predicted.
In the $B=2$ sector there are many predictions, but only few of them - e.g.,
the virtual $\Lambda N$ state - have been observed. Possible reasons for
this are discussed.
\vglue 0.4cm
\elevenrm
\baselineskip=13pt
\section{\elevenbf INTRODUCTION. THE $B=1$ SECTOR}
 There is growing evidence now that the chiral soliton approach (CSA) 
proposed
at first by Skyrme allows one to describe the properties of baryons - octet
and decuplet \cite{1} - and also the basic properties of lightest nuclei with 
baryon number $B=2,3$ and $4$ \cite{2}.
This approach has rich consequences
for the spectra of baryons and baryonic systems with nonzero strangeness.
However, its status in the sectors with baryon number $B=1$ and $B \geq 2$ is 
quite different.
In the $B=1$ sector the octet and decuplet of baryons are known for many
years. When the chiral soliton approach was applied for description of
the baryon properties it was a problem to describe corresponding mass 
splittings \cite{3}. Reasonable agreement with data has been reached when 
the flavor symmetry breaking was taken into account not only in meson masses 
but also in the meson decay constants, with $F_K/F_{\pi} \simeq 1.26$ \cite{1}.

Within the CSA the mass formula for the quantized baryonic states has the
following general structure:
$$ M = M_{cl} + E_{rot}(p,q,Y,T,J; \Theta_T, \Theta_S, \Theta_J,...) + E_{Cas}
 \eqno (1) $$
where $M_{cl}$ is the classical mass of the soliton with baryon number $B$
including the symmetry breaking terms, $M_{cl} \sim N_c$, the number of colors
in the underlying QCD. $E_{rot}$ is the zero-modes quantum correction 
depending on the quantum numbers of state and moments of inertia $\Theta_i$
defined by the profile functions of the solitons. $E_{rot}$ contains the
terms $\sim 1/N_c$ and terms $\sim N_c^0$, as it was shown recently by
Walliser \cite{4}. $p,q, Y$, $T$ and $J$ denote 
the $SU(3)$ multiplet, hypercharge of the state, its isospin and angular
momentum.
$E_{Cas} \sim N_c^0$ is the
so called Casimir energy of solitons closely connected with the loop 
corrections to the mass of solitons. It is a problem to calculate this energy
because the model is not renormalizable. Till now $E_{Cas}$ was estimated 
in few cases only \cite{5,6,4}.

 In the simplest case of $B=1$ only two moments of inertia come into the game,
$\Theta_T=\Theta_J$ and $\Theta_S$, so called strange, or kaonic inertia.
The rotational energy depending on $\Theta_S$ for a state with any $B$-number 
equals to ($N_c=3$) :
$$E_{rot}(\Theta_S,B) = [3B/2 + m(3B/2+m+1-N)]/ (2\Theta_S (B))  \eqno (2) $$
Here $m$ is the difference between triality and the $B$-number,
$m=(p+2q)/3 - B$ which can be interpreted as the number of additional quark-
antiquark pairs present in the multiplet $(p,q)$, $N=(p+m)/2$ is 
the "right" isospin.
For "minimal" $SU(3)$ multiplets $m=0$. For $B=1$ octet and decuplet of baryons
are just such minimal multiplets, $(p+2q)/3=1$. The first term
in $(2)$, $3B/(4\
Theta_S)$, is the same for all minimal multiplets because of 
cancellation of the second order Casimir operators of the groups $SU(2)$ and
$SU(3)$ \cite{7}.
The antidecuplet and $27$-plet of baryons contain exotic states which cannot
be constructed from the valence quarks only. The additional
energy of these multiplets depending on the inertia of baryon  
 $\Theta_S (1)$ is given by second term in $(2)$ proportional to $m$, 
with $m=1, B=1$ \cite{7}.

In many variants of the model for baryons $\Theta_S(1) \simeq 2.1 Gev^{-1}$
\cite{1,8},
therefore, for $m=1, N=1/2$ (antidecuplet) the term $\sim m$ in $(2)$
$\Delta E_{rot,\bar{10}} = 3/(2\Theta_S) \simeq 714 Mev$.
Antidecuplet contains the positive strangeness, $S=+1$, isospin $T=0$ 
component $Z^+$ which is of special interest as the lowest baryon with
positive strangeness. 
The first numerical calculation of the mass difference of $Z^+$ and nucleon
gave $\Delta M_{Z^+}=M_{Z^+} - M_N \simeq 740 Mev$ \cite{8}. The 
Schwesinger-Weigel slow rotator approach fitting the mass splittings of 
the octet and decuplet of baryons
gives $\Delta M_{Z^+} = 795 Mev$. Recently the problem of antidecuplet
received much attention in \cite{9}. Fitting the mass difference of $N^*(1710)$
resonance and the nucleon in the assumption that $N^*(1710)$ is just the
nonstrange component of $\bar{10}$ the authors obtained 
$\Delta M_{Z^+} = 590 Mev$
and estimated the width of $Z^+$ to be $\sim 15 Mev$. The searches of $Z^+$
as well as exotic $\Xi^{*--}$ and $\Xi^{*+}$ states, the components of the 
iso-quartet with $S=-2$, are of interest. The latter state should have the
mass about $1.2 Gev$ greater than nucleon mass and the width of several tens
of $Mev$, at least.
\section{\elevenbf THE $B=2$ SECTOR: $SO(3)$ HEDGEHOG AND $SU(2)$ TORUS}
Situation is different in the sector with baryon number $B=2$.
The chiral soliton approach is of special
interest for $B \geq 2$ because it provides unconventional point of view at the
baryonic systems and nuclear fragments. The baryons individuality is
absent in the bound states of skyrmions and can be restored when non-zero
modes quantum effects are taken into account \cite{2}, \cite{10}.

Till now three different types of dibaryons are established within
the chiral soliton approach.
There is longstanding prediction of the $H$-dibaryon in the framework
of MIT quark-bag model \cite{11}, confirmed  also 
in the $SU(3)$ extension of the Skyrme model \cite{12} and in the quark-
cluster model \cite{13}. The $SO(3)$ hedgehog
with the lowest possible for this subgroup baryon (winding) number, $B=2$, 
is interpreted usually as $H$-dibaryon. Big efforts of different
groups during several years did not lead to the experimental confirmation of 
this prediction till present time.

The state with the azimuthal winding $n=2$ has $B=4$ and the torus-like 
form of the mass and $B$-
number distributions. It is bound relative to the 
decay into two $B=2$ hedgehogs \cite{14}. This tendency of binding of two
$H$-particles has been confirmed also within some variant of the quark model
\cite{15}.

As it was shown in \cite{6} the $H$-particle can be unbound when 
Casimir energies (CE) of solitons are taken into account. It should be
noted that, within the CSA, $H$-particle is an object considerably smaller
than the deuteron, $<R_H>^2 \simeq (0.2 - 0.3) Fm^2$, \cite{6}. Therefore, 
theoretical estimates of the $H$ production cross sections based on the 
similarity
between $H$ and the deuteron can be overestimated at least by one order of
magnitude.

The second type of dibaryons is obtained by means 
of the quantization of bound $SU(2)$-solitons in $SU(3)$ collective 
coordinates space. The bound state
of skyrmions with $B=2$ possesses generalized axial symmetry and
torus-like distributions of the mass and $B-$number densities \cite{16}. Now
it is checked in several variants of chiral soliton models and
also in the chiral quark-meson model. Therefore, the existence
of $B=2$ torus-like bound skyrmion seems to be firmly established. 

After zero-modes quantization procedure the $SU(3)$ multiplets of dibaryons
appear, with the ratio of strangeness to baryon number 
$S/B$ down to $-3$. The possible $SU(3)$ multiplets which could consist
of minimal number of valence quarks are antidecuplet, $27-$, $35-$ and 
$28-$ plets. The contribution to the energy from rotations
into "strange" direction is the same for minimal irreps satisfying
the relation $\frac{p+2q}{3}=B=2$, as it was noted above, see Eq. (2). 
The deuteron
binding energy within this approach is about $30 Mev$ instead of $2.2 Mev$,
so, $ \sim 30 Mev$ is uncertainty of the model predictions. All strange
states are bound when contributions linear in $N_c$, the classical
mass, and also $E_{rot}$ are taken into account. However,
after renormalization of masses which is necessary to take into 
account also the CE of the torus (of the order $N_C^0$) and to 
produce the nucleon-nucleon $^1S_0$-scattering state on the right 
place all states with strangeness
different from zero are above thresholds for the strong decays \cite{17}.
Therefore, it will be difficult to observe such states experimentally. 
The virtual $\Lambda N$ level seen many years ago in reaction $pp \rightarrow
p\Lambda K^+$ \cite{18} and confirmed in recent measurements can be one of
the states with $S=-1$ obtained in \cite{17}. 
\section{\elevenbf SKYRMION MOLECULES} 
 The third type of states
is obtained by means of quantization of strange skyrmion molecules (SSM)
found recently \cite{19}.
To obtain the strange skyrmion molecule we used the ansatz of the
type
$$        U = U_L(u,s)  U(u,d)  U_R(d,s)            \eqno   (3) $$
where  $U_L(u,s)$ and $U_R(d,s)$ describe solitons located in $(u,s)$ 
and $(d,s)$ $SU(2)$ subgroups of $SU(3)$, one of $SU(2)$-matrices, 
 e.g. $U(u,d)$   depends  on two parameters:
$$    U(u,d)= \exp(ia \lambda_2 ) \exp(ib \lambda_3 ) \eqno (4) $$ 
and thus describes the relative local orientation
of these solitons in usual isospace. The configuration considered
depends totally on 8 functions of 3 variables. 
It should be noted that the baryon number density as well as chirally
invariant contributions to the energy of solitons can be presented in the
form symmetric in different $SU(2)$ subgroups of $SU(3)$:
$$B=-{1\over 2\pi^2}\int \bigl[ (\vec{L}_1\vec{L}_2\vec{L}_3)+(\vec{L}_4
\vec{L}_5\tilde{\vec{L}}_3)+(\vec{L}_6\vec{L}_7\tilde{\tilde{\vec{L}}}_3) 
+{1\over 2}[(\vec{L}_1,
\vec{L}_4\vec{L}_7-\vec{L}_5\vec{L}_6)+(\vec{L}_2,\vec{L}_4\vec{L}_6+
\vec{L}_5\vec{L}_7) ] \bigr] d^3r \eqno (5) $$ 
Here $\tilde{L}_3=(L_3+\sqrt{3} L_8)/2$, $\tilde{\tilde{L}}_3=(-L_3
+\sqrt{3} L_8)/2$ are the $3$-d components of chiral derivatives in the
$(u,s)$ and $(d,s)$ $SU(2)$ subgroups of $SU(3)$, $i\lambda_k\vec{L}_k=
U^{\dagger}\vec{\partial} U$, $\lambda_k$ are $8$ Gell-Mann matrices.
 
To get the $B=2$ molecule we started from two $B=1$ skyrmions in
the optimal attractive orientation at relative distance between 
topological centers close to the optimal one, a bit smaller.
Special algorithm
for minimization of the energy functionals depending on 8 functions
was developed and used \cite{19}. 
The energy functional of arbitrary $SU(3)$ solitons can be written also in
the form which respects the democracy of different $SU(2)$ subgroups of 
$SU(3)$ \cite{20}:
$$M_{cl}=\int \bigl(M_2 +M_4 +M_{SB}\bigr) d^3r \eqno (6) $$
With
$$M_2= \frac{F_{\pi}^2}{8} \bigl[\vec{L}_1^2+\vec{L}_2^2+\vec{L}_4^2+
\vec{L}_5^2+\vec{L}_6^2+\vec{L}_7^2+
{2 \over 3}(\vec{L}_3^2+
\tilde{\vec{L}}_3^2+\tilde{\tilde{\vec{L}}}_3^2)\bigr] \eqno (7) $$
$$M_4={1 \over 4e^2} \Biggl\{ (\vec{s}_{12}+\vec{s}_{45})^2+(\vec{s}_{45}+
\vec{s}_{67})^2+(\vec{s}_{67}-\vec{s}_{12})^2+{1 \over 2} \bigl[
(2\vec{s}_{13}-\vec{s}_{46}-\vec{s}_{57})^2+(2\vec{s}_{23}+\vec{s}_{47}-
 \vec{s}_{56})^2+ $$
$$+(2\tilde{\vec{s}}_{34}+\vec{s}_{16}-\vec{s}_{27})^2+(2\tilde{\vec{s}}_{35}
+\vec{s}_{17}+\vec{s}_{26})^2+
(2 \tilde{\vec{s}}_{36}+\vec{s}_{14}+\vec{s}_{25})^2+(2\tilde{\vec{s}}_{37}+
\vec{s}_{15}-\vec{s}_{24})^2 \bigr] \Biggr\} \eqno (8) $$
$\vec{s}_{ik}=[\vec{L}_i\vec{L}_k]$, 
$\tilde{\vec{s}}_{34}=[\tilde{\vec{L}}_3\vec{L}_4]$, and 
$\tilde{\vec{s}}_{36}=[\tilde{\tilde{\vec{L}}}_3 \vec{L}_6]$,
similar for $\tilde{\vec{s}}_{37}$. Note that $\vec{L}_8$ or 
$\tilde{\vec{L}}_8$ do not enter $(5)-(8)$. 

The mass term $M_{SB}$ violates the chiral symmetry and contains the flavor
symmetric as well as flavor symmetry breaking parts:
$$M_{SB}=F_{\pi}^2m_{\pi}^2(3-v_1-v_2-v_3)/8 +(F_K^2m_K^2-F_{\pi}^2m_{\pi}^2)
 (1-v_3)/4, \eqno (9) $$
$v_1,v_2,v_3$ are the real parts of the diagonal elements of $SU(3)$ matrix 
$U$.
Expressions $(5)-(9)$ and $(10)$ below provide the framework for studies of 
any $SU(3)$ skyrmions located originally in arbitrary $SU(2)$ subgroups 
of $SU(3)$.

After minimization of the energy functional we obtained the
configuration of molecular type with the binding energy about half
of that of the torus, i.e. about $\sim 75$ $Mev$ for parameters of the
model $F_{\pi}=186$ $Mev$ and $e=4.12$ \cite{19}. The attraction between unit 
skyrmions which led to the formation of torus-like configuration
when they were located in the same $SU(2)$ subgroup of $SU(3)$
is not sufficient for this when solitons are located in different
subgroups of $SU(3)$. It is connected with the fact that solitons
located in different $SU(2)$ subgroups interact through one common
degree of freedom, instead of $3$ degrees, as in the first case.
\section{\elevenbf QUANTIZATION OF THE SSM}
 The quantization of zero modes of strange skyrmion molecules cannot
be done using the standard procedure, its substantial modification
is necessary \cite{20}. To proceed we calculated the Wess-Zumino term
for arbitrary $SU(3)$ skyrmions. It is linear in the angular velocities
of rotation in the $SU(3)$ configuration space defined in the standard way:
$A^{\dagger} \dot{A}=-{i \over 2} \omega_k\lambda_k$, $k=1,...8$, and
$WZ \sim (WZ^L_k+WZ^R_k)\omega_k$. 
The $8$-th component of the WZ-term is most important and is equal to
$$WZ_8^L=-\sqrt{3}(\vec{L}_1\vec{L}_2\vec{L}_3) +(\vec{L}_8\vec{L}_4\vec{L}_5)
+(\vec{L}_8\vec{L}_6\vec{L}_7) \eqno(10) $$
Similar expression holds for $WZ_R$ in terms of right chiral derivatives
$\vec{R}_k$.
As a result, the quantization condition
established first in \cite{3} is changed, and for strange
skyrmion molecule we obtained
$$ Y^{min}_R = 2 \partial L^{WZ}/(\sqrt{3} \partial \omega_8) \simeq 
-(B_L+B_R)/2 = -1  \eqno(11)  $$
for $B_L=B_R=1$, \cite{20} instead of the known relation for the right
hypercharge $Y_R=B$, \cite{3} (we put here the number
of colors $N_c=3$), $B_L$ and $B_R$ are the $B$-numbers located in $(u,s)$ and
$(d,s)$ $SU(2)$ subgroups.
The interpolating formula proposed in \cite{19} for $Y^{min}_R$ is
$$Y_R^{min} \simeq N_cB(1-3C_S)/3 \eqno(12) $$
with $C_S=<1-v_3>/<3-v_1-v_2-v_3>$ - scalar strangeness content of solitons.
$(12)$ is exact for any $(u,d)$ $SU(2)$ solitons rotated in $SU(3)$ 
collective coordinates space as well as for $SO(3)$ solitons ($C_S=1/3$).
 For strange 
 molecule $C_S \simeq 1/2$ and $(12)$ is valid approximately \cite{20}. 

The zero-modes energy - quadratic form in $8$ angular velocities of rotation
in $SU(3)$ configuration space - can be obtained from $(8)$ by means of 
substitution in $M_2$ $\vec{L}_i \rightarrow \tilde{\omega}_i/2$, and in
$M_4$ $\vec{s}_{ik} \rightarrow [\tilde{\omega}_i\vec{L}_k-\tilde{\omega}_k
\vec{L}_i]/2$, $\tilde{\omega}_i$ being some linear combination of 8 components
of angular velocities $\omega_i$, details can be found in \cite{20}.
The moments of inertia of $SU(3)$ skyrmions can be calculated from this
expression.

In view of the evident relation $(p+2q)/3 \geq Y^{min}_R \geq -(q+2p)/3 $ the 
lowest multiplets obtained by means of quantization of
strange skyrmion molecule are octet, decuplet and antidecuplet
with central values of masses about $4.2$, $4.5$ and $4.7$ $Gev$.
Within the octet the states with strangeness $S=-1$, $-2$ and $-3$
are predicted. They are coupled correspondingly to $\Lambda N$-$\Sigma N$, 
$\Lambda\Lambda$-$\Xi N$
or $\Lambda\Sigma$ and $\Lambda\Xi$-$\Sigma\Xi$ channels, see the discussion
of the absolute values of masses within CSA below in Section 5.

The mass splittings within multiplets considered are defined, as
usually, by chiral and flavor symmetry breaking mass term in the 
effective lagrangian. Its contribution to the masses of the states 
in the case of strange skyrmion molecules equals to
$$\delta M = - \frac{1}{4} (F^2_Km^2_K-F^2_{\pi}m^2_{\pi})
 (v_1+v_2-2v_3) <sin^2\nu/2>    \eqno(13) $$
The function $\nu$ parametrizes as usually the 
$\lambda_4$ rotation in the collective coordinates quantization 
procedure, and the
average over the wave function of the state should be taken for
$sin^2\nu$. For two interacting undeformed hedgehogs at large
relative distances $v_1+v_2-2v_3 \rightarrow 2(1-cosF)$ where $F$
is the profile function of the hedgehog. Note, that the sign in
$(13)$ is opposite to the sign of analogous term when $(u,d)$
$SU(2)$ soliton is quantized with $SU(3)$ collective coordinates.

The result of calculation depends to some degree on the way of
calculation. We can start with the soliton calculated for all
meson masses equal to the pion mass (flavor symmetric, FS-case),
or with soliton calculated with the kaon mass included into the
lagrangian (FSB-case). The static energy of solitons are greater
in the FSB case, the moments of inertia are smaller, and the mass
splittings within the $SU(3)$ multiplets are squeezed by a factor
about $\sim 2.5$ in the latter case in comparison with the FS-case \cite{20}.
 The results of both ways of calculation are close to each other
for the octet of dibaryons, the difference increases for decuplet
and is large for antidecuplet. By this reason the method of
calculation should be found where results do not depend on the
starting configuration. It can be, probably, some kind of 
"slow rotator" approximation \cite{1}.

The relative binding energy of quantized states ranges from $\sim 0.14$ for
the octet, to $\sim 0.11$ for decuplet down to $\sim 0.07$ for antidecuplet
of dibaryons.

The inclusion of the configuration mixing into consideration \cite{21}
usually increases the mass splittings within multiplets, although it
does not change the results crucially.
\section{\elevenbf SUMMARY AND DISCUSSION}
 To summarize, there are different branches of the predictions of strange
dibaryons within the chiral soliton approach. The first one is 
the $SO(3)$ hedgehog identified usually with the $H$-particle predicted within
the MIT quark-bag model. The second is obtained
by means of the quantization of the bound torus-like biskyrmion.
The third is the quantized strange skyrmion molecule.

The main uncertainty in the masses of all predicted states comes from
the poor known Casimir energies of states - the loop corrections 
of the order of $N_c^0$ to the classical masses of solitons. 
The CE was estimated for
the $B=1$ hedgehog \cite{5,4} and also for $B=2$ $SO(3)$ hedgehog
\cite{6}. For $B=1$ case it has right sign and order of magnitude,
about $(-1  - -1.5)$ $Gev$. For the torus-like
$B=2$ skyrmion \cite{16} the CE is not estimated yet.

The skyrmion molecules found in \cite{19} should have the lowest 
uncertainty in Casimir energies relative to the $B=1$ states since 
in the molecule
unit skyrmions are only slightly deformed in comparison with starting
unperturbed configurations. Therefore, one can hope that the property of 
binding of dibaryons belonging to the lowest multiplets, octet and
decuplet, will not dissappear after inclusion of the CE and vibration,
breathing, etc. quantum corrections.
The results for strange molecule are in qualitative agreement with those
of \cite{22} where the attraction between hyperons was found at large
relative distances.

The prediction of the existence of multiplets of strange dibaryons
 some of them being bound relative to strong 
interaction, remains a challenging property of the chiral soliton 
approach. Quite similar predictions can be obtained also for baryonic
systems with $B=3,4$, etc. These predictions are on the same level as the 
existence of strange hyperons in the $B=1$ sector of the model because, 
within the
chiral soliton approach, skyrmions with different values of $B$ are
considered on equal footing. Further
theoretical studies and comparison with predictions of other models
(see, e.g. \cite{23}-\cite{25}) would be important, as well as the experimental 
searches for such states. The enhancement of strangeness production observed
in heavy ion collisions can be, at least partly, due to copious production and
subsequent decays of strange baryonic systems (nuclear fragments).

However, in view of specific internal problems of the CSA - e.g., both $SO(3)$
hedgehog and strange $B=2$ molecule do not possess definite parity \cite{12,20}
 - it can be that some of the predictions of the Skyrme model are 
the artefact of the model. If it is really so, one should understand the 
reason for this, and how to separate true prediction from the wrong one.

I am indebted to H.Walliser for useful discussions, remarks 
and for sending me the program for the configuration mixing 
calculations, and also to D.J.Millener, B.E.Stern for their help.
 


\newpage
{\elevenbf\noindent References}
\begin{thebibliography}{29}
\baselineskip=12pt
\bibitem{1} B. Schwesinger, H. Weigel, Phys. Lett. B267(1991)438;
H. Weigel, Int. J. Mod. Phys. A11(1996)2419 
\bibitem{2} E. Braaten, L. Carson, Phys.Rev. D38(1988)3525;
 L. Carson, Nucl. Phys. \\
A535(1991)479; T. Walhout, Nucl. Phys. A547(1992)423; T. Waindzoch,\\
 J. Wambach, Phys. Lett. B226(1992)163;
R. A. Leese, N. Manton, B. J. Schroers,\\
 Nucl. Phys. B442(1995)228; N. R. Walet, Nucl. Phys. A586(1995)649; \\
Nucl. Phys. A606(1996)429 (hep-ph/9603273)
\bibitem{3} E. Guadagnini, Nucl. Phys. B236(1984)35; M. Chemtob, Nucl. Phys.
B256(1985)600; M. Praszalowicz, Phys. Lett. 158B(1985)264
\bibitem{4} H. Walliser, hep-ph/9710232
\bibitem{5} B. Moussalam, Ann. of Phys. (NY) 225(1993)264; G. Holzwarth,
H. Walliser,\\ Nucl. Phys. A587(1995)721; F. Meier, H. Walliser, Phys.Rept. 
289(1997)383 \\
(hep-ph/9602359)
\bibitem{6} F.G. Scholtz, B. Schwesinger, H.B. Geyer, Nucl. 
Phys. A561(1993)542  
\bibitem{7} V. B. Kopeliovich, Phys. Lett. 259B(1991)234; Nucl. 
Phys. A547(1992)315c
\bibitem{8} H. Walliser, Nucl. Phys. A548(1992)649
\bibitem{9} D. Diakonov, V. Petrov, M. Polyakov, hep-ph/9703373 
\bibitem{10} V.B. Kopeliovich, Phys. Atom. Nucl. 56(1993)1084; 
Yad. Fiz. 47(1988)1495
\bibitem{11} R.L. Jaffe, Phys. Rev. Lett. 38(1977)195 
\bibitem{12} A.P. Balachandran et al., Phys. Rev. Lett. 52(1984)887; 
Nucl. Phys. B256(1985)525; R. L. Jaffe, C. L. Korpa, Nucl. 
Phys. B258(1985)468
\bibitem{13} Y. Koike, K. Shimizu, K. Yazaki, Nucl. Phys. A513(1990)653
\bibitem{14} A.I. Issinskii, V.B. Kopeliovich, B.E. Stern, 
Sov. J. Nucl. Phys. 48(1988)133 \\
(Yad.Fiz. 48(1988)209)
\bibitem{15} T. Sakai, J. Mori, A.J. Buchmann,
K. Shimizu, K. Yazaki, nucl-th/9709054
\bibitem{16} V.B. Kopeliovich, B.E. Stern, JETP Lett. 45(1987)203; 
   J.J.M. Verbaarschot, \\Phys. Lett. 195B(1987)235;
N. S. Manton, Phys. Lett. 192B(1987)177
\bibitem{17} V. Kopeliovich, B. Schwesinger, B. Stern, Nucl. 
Phys. A549(1992)485
\bibitem{18} J.T. Reed et al, Phys. Rev. 168(1969)1495; W.G.
Hogan et al, Phys.Rev. \\
166(1968)1472
\bibitem{19} V.B. Kopeliovich, B.E. Schwesinger, B.E. Stern,
 JETP Lett. 62(1995)185 (Pis'ma v ZhETF 62(1995)177)
\bibitem{20} V. Kopeliovich, ZhETF 112(1997)1241 (hep-th/9707067); 
JETP Lett. 64(1996)426
\bibitem{21} H. Yabu, K. Ando, Nucl. Phys. B301(1988)601
\bibitem{22} B. Schwesinger, F.G. Scholtz, H.B. Geyer, Phys. 
Rev. D51(1995)1228
\bibitem{23} T. Goldman et al, Phys. Rev. Lett. 59(1987)627;
F. Wang,J.-l. Ping,G.-h.Wu, \\L.-j. Teng,T. Goldman, Phys. 
Rev. C51(1995)3411 (nucl-th/9512014)
\bibitem{24} A. Gal, C.B. Dover, Nucl. Phys. A585(1995)1c; C.B. Dover, A.
Gal, Nucl. Phys. \\
A560(1993)559; D.J. Millener, C.B. Dover, A. Gal, Phys. 
Rev. C38(1988)2700
\bibitem{25} E. Farhi, R. Jaffe, Phys. Rev. D30(1984)2379 \\
H. Heiselberg, C.J. Pethick, Phys. Rev. Lett. 70(1993)1355
\end{thebibliography}
\end{document}